p

# Are there laws of genome evolution?

Eugene V. Koonin

National Center for Biotechnology Information, National Library of
Medicine, National Institutes of Health, Bethesda MD, USA

p

**Abstract**

Research in quantitative evolutionary genomics and systems biology led to the discovery of several universal regularities connecting genomic and molecular phenomic variables. These universals include the log-normal distribution of the evolutionary rates of orthologous genes; the power law-like distributions of paralogous family size and node degree in various biological networks; the negative correlation between a gene's sequence evolution rate and expression level; and differential scaling of functional classes of genes with genome size. The universals of genome evolution can be accounted for by simple mathematical models similar to those used in statistical physics, such as the birth-death-innovation model. These models do not explicitly incorporate selection, therefore the observed universal regularities do not appear to be shaped by selection but rather are emergent properties of gene ensembles. Although a complete physical theory of evolutionary biology is inconceivable, the universals of genome evolution might qualify as 'laws of evolutionary genomics' in the same sense 'law' is understood in modern physics.

p

**Author Summary**

Research in comparative genomics and systems biology uncovered several universal, quantitative regularities of genome evolution, such as the distribution of the evolutionary rates of orthologous genes that is virtually indistinguishable from bacteria to mammmals, and anti-correlation between a gene's sequence evolution rate and expression level. What are these universals of genome evolution? Should they be considered 'laws of genome evolution' or biologically irrelevant statistical effects? Here I discuss simple mathematical models similar to those used in statistical physics that can account for the universals of genome evolution and argue that, although a complete physical theory of evolutionary biology is inconceivable, these universals might qualify for the status as physical laws.

p

**Introduction**

Darwin's concept of evolution, all its generality and plausibility notwithstanding, was purely qualitative. In the 1920s and 1930s, seminal work of Fisher, Wright and Haldane laid the foundation for quantitative analysis of elementary processes in evolving population, and in the 1950s, this population genetic theory was incorporated in the framework of the Modern Synthesis of evolutionary biology. However, the formalism of population applies only to microevolution in idealized populations and falls far short of a general quantitative theory of evolution. Rapid progress of genomics and systems biology at the end of the $20^{th}$ century and in the beginning of the $21^{st}$ century brought about enormous amounts of new data amenable to quantitative analysis. The new data types include numerous complete genome sequences, transcriptomes (genome-wide gene expression information), proteomes (organism-wide protein abundance information), interactomes (organism-wide data on physical and genetic interactions between proteins or gene), regulomes (comprehensive data on gene expression regulation) and more. This deluge of new information spawned a research direction that occupies itself with quantification of the relationships between various genomic and molecular phenomic variables and may be called quantitative evolutionary genomics [1,2].

**Universals of genome and molecular phenome evolution**

Quantitative comparative genomic analysis revealed several universals of genome evolution that come in the form of distinct distributions of certain quantities or specific dependencies between them. The most conspicuous universals include (Figures 1 and 2):

p

- log-normal distribution of the evolutionary rates between orthologous genes[3,4,5];
- power law-like distributions of membership in paralogous gene families and node degree in biological 'scalefree' networks[6,7,8,9];
- negative correlation between a gene's sequence evolution rate and expression level (or protein abundance)[10,11,12,13];
- distinct scaling of functional classes of genes with genome size[14,15].

The universality of these dependencies appears genuinely surprising. For example, the distributions of sequence evolution rate of orthologous genes are virtually indistinguishable in all evolutionary lineages, for which genomic data are available, including diverse groups of bacteria, archaea and eukaryotes[3,4,5]. The shape of the distribution did not perceptibly change through about 3.5 billion years of the evolution of life even though the number of genes in the compared organisms differs by more than an order of magnitude, and the repertoires of gene functions are dramatically different as well[5]. The same conundrum pertains to the other universals: despite major biological differences between organisms, these quantitative regularities hold, often to a high precision. What is the nature of the genomic universals? Do they reflect fundamental 'laws' of genome evolution or are they 'just' pervasive statistical patterns that do not really help us understand biology? A related major question is: are these universals affected or maintained by selection?

p

**Mathematical models to account for the evolutionary universals**

Clearly, should there be laws of genome evolution, in the sense this term is used in physics, identification of recurrent patterns and universal regularities is only the first step in deciphering these laws. The obvious next steps involve developing physical (mathematical) models of the evolutionary processes that generate the universals and test the compatibility of the predictions of these models with the observations of comparative genomics and systems biology. Indeed, such models have been proposed to account for each of the universals listed above (Figure 2). Notably, these models can be extremely simple, based on a small number of biologically plausible elementary processes, but they are also highly constrained. A case in point is the birth-death-and-innovation model that explains the power law-like distribution of gene family sizes in all genomes[7,8,9]. This model includes only three elementary processes, the biological relevance of which is indisputable: i) gene birth (duplication), ii) gene death (elimination), iii) innovation (that is, acquisition of a new family, e.g., via horizontal gene transfer). A model with precise balance between the rates of these elementary processes and a particular dependency of birth and death rates on paralogous family size yields family membership distributions that are statistically indistinguishable from the empirically observed distributions[7].

Straightforward models of evolution have been developed that apparently account for more than one universal (Figure 2). A case in point is a recent amended birth-death-innovation model of evolution that connects two genomic universals that are not obviously related, namely, the distribution of gene family size and differential scaling of functional classes of genes with the genome size[16]. In this model, gain and loss rates of

p

genes in different functional classes (e.g., metabolic enzymes and expression regulators) are linked in a biologically motivated proportion. The model jointly reproduces the power-law distribution of gene family sizes and the non-linear scaling of the number of genes in functional classes with genome size. Moreover, the model predicted that functional classes of genes that grow faster-than-linearly with genome size would show flatter-than-average family size distributions. The existence of such a link between these *a priori* unrelated exponents is indeed confirmed by analysis of prokaryotic genomes.

The ubiquitous negative correlation between sequence evolution rate and expression level triggered the hypothesis of misfolding-driven protein evolution that explains the universal dependency between evolution and expression under the assumption that protein misfolding is the principal source of cost incurred by mutations and errors of translation[4,17]. This assumption was used to incorporate evolutionary dynamics into an off-lattice model of protein folding[18]. The resulting model of protein evolution reproduced, with considerable accuracy, the universal distribution of protein evolutionary rates, as well as the dependency between evolutionary rate and expression. These findings suggest that both universals of evolutionary genomics could be direct consequences of the fundamental physics of protein folding.

**Universals of evolution are emergent properties of gene ensembles not selectable features**

The models of evolution that generate the observed universal patterns of genome evolution do not explicitly incorporate selection. The question of selective vs neutral

p

emergence of global quantitative regularities has been explored in some detail for the case of network architectures. Networks have become ubiquitous images and tools of systems biology[6]. Indeed, any class of interacting objects can be naturally represented by nodes, and the interactions between these objects, regardless of their specific nature, can be represented by edges. Commonly explored biological networks represent gene coexpression; genetic interactions between genes; physical interactions between proteins; regulatory interactions between genes; metabolic pathways where metabolites are nodes and enzymes are associated with edges; and more, considering that the network formalism is general and flexible enough to capture all kinds of relationships. In a sharp contrast to random networks that are characterized by a Poisson distribution of the node degree, biological networks typically show a power-law-like node degree distribution, $P(k) \sim k^{-\gamma}$, where $k$ is the node degree, i.e., the number of nodes to which the given node is connected and $\gamma$ is a positive coefficient. These networks are said to be scale-free because the shape of their node degree distribution remains the same regardless of the chosen scale, that is any subnetwork is topologically similar to the complete network (in other words, scale-free networks display fractal properties). The negative power law node degree distribution is characteristic not only of biological networks but also of certain purely "artificial" networks such as the Internet. Barabasi and colleagues came up with the provocative idea that this is an intrinsic feature of evolved networks and proposed a simple and plausible mechanism of network evolution known as preferential attachment[19]. In addition to the scale-free architecture, most of the biological networks possess additional interesting features such as small world properties, modularity and

p

hierarchical structure that are also widespread but tend to differ among networks representing different classes of biological phenomena[6].

Scale-free networks are "robust to error but vulnerable to attack": elimination of a randomly chosen node most of the time has little effect on the overall topology and stability of the network whereas elimination of highly connected nodes (hubs) disrupts the network. This property might be conceived as implying that the architecture of such networks represents "design" that evolved under selection for increased robustness. However, this idea is no more justified than the view that the Internet was deliberately designed with the same purpose in mind. The preferential attachment mechanism in itself is a non-adaptive route of network evolution. Simulation of the growth of a network by random duplication of its nodes with all their connections followed by subfunctionalization, i.e., differential loss of edges by the daughter nodes not only yields the typical power law distribution of the node degree but also reproduces the modular structure of biological (specifically, protein-protein interaction) networks[20]. Duplication followed by subfunctionalization is the most common route of gene evolution that does not intrinsically involve selection. Rather, subfunctionalization is naturally interpreted as a type of "constructive neutral evolution" whereby complexity, and complex networks in particular, evolve not as adaptations but through irreversible emergence of dependencies between parts of the evolving system[21,22].

Compelling evidence of the non-adaptive origin of global architectural features of networks was obtained through the analysis of gene coexpression networks in mutation accumulation (MA) lines of the nematode *Caenorhabditis elegans*[23]. The MA lines are

p

virtually free of selective constraints, so comparison between these lines and natural isolates provides for evaluation of the contribution of selection to the evolution of various characters, in particular network architecture. The global architectures of evolutionary coexpression networks (i.e., networks in which edges connected genes with similar patterns of expression across multiple lines) were indistinguishable between MA lines and natural isolates, demonstrating that these features are not subject to selection. Furthermore, there was no significant correlation between the properties of any given node, such as the degree and the clustering coefficient, in the networks from mutation accumulation lines and natural isolates. These results strongly suggest that not only general architectural properties of networks but even the position of individual nodes in networks are not subject to substantial selection.

Collectively, the ability of simple models to generate the universals of genome evolution and additional results indicating that the global architecture of biological networks is not a selected feature suggest that all evolutionary universals are not results of adaptive evolution. Such a conclusion does not imply that these universals are biologically irrelevant: beneficial properties such as network robustness may emerge "for free" from the most general principles of evolution.

The universal dependencies and distributions seem to be emergent properties of biological systems that appear because these systems consist of numerous (sufficiently numerous for the manifestation of robust statistical regularities) elements (genes or proteins, depending on the context) that weakly interact with each other, compared to the strong interactions that maintain the integrity of each element. Clearly, this

p

representation of biological systems as ensembles of weakly interacting "particles" resembles rough but enormously useful approximations, such as ideal gas, that are routinely used in statistical physics. This approach is obviously over-simplified because higher level interactions such as epistasis are common and critically important in biology[24,25]. Nevertheless, the ability of simple models akin to those used in statistical physics to quantitatively reproduce universals of genome and molecular phenome evolution attest to the fruitfulness of the "statistical ensemble" approximation.

### 'Laws' of evolutionary genomics

The analogies between the evolutionary process and statistical physics are not limited to the existence of universal dependencies and distributions, some of which can be derived from simple models. It is actually possible to draw a detailed correspondence between the key variables in the two areas [26,27]. The state variables (degrees of freedom) in statistical physics such as positions and velocities of particles in a gas are analogous either to the states of sites in a nucleotide or protein sequence, or to the gene states in a genome, depending on the level of evolutionary modeling. The characteristic evolutionary rate of a site or a gene naturally corresponds to a particle velocity. Furthermore, effective population size plays a role in evolution that is clearly analogous to the role of temperature in statistical physics, and fitness is a natural counterpart to free energy.

The process and course of evolution critically depend on historical contingency and involve extensive adaptive "tinkering"[28,29]. Therefore a complete physical theory of evolution (or any other process with a substantial historical component) is inconceivable.

Nevertheless, the universality of several simple patterns of genome and molecular phenome evolution, and the ability of simple mathematical models to explain these universals suggest that "laws of evolutionary biology" comparable in status to laws of physics might be attainable.

p

Figure legends

Figure 1. Universals of genome and molecular phenome evolution

The figure shows idealized versions of universal dependencies and distributions. The scattered points show the range of characteristic variance.

- (A) Log-normal distribution of evolutionary rates of orthologous genes
- (B) Anticorrelation between gene expression level (protein abundance) and sequence evolution rate
- (C) Power law-like distribution of paralogous family size
- (D) Differential scaling of functional classes of genes with the total number of genes in a genome. Three fundamental exponents are thought to exist: 0 – no dependence, typical of translation system component; 1 – linear dependence, characteristic of metabolic enzymes; 2 – quadratic dependence, characteristic of regulatory and signal transduction system components.

Figure 2. Universals of genome and molecular phenome evolution and underlying physical/mathematical models.

Arrows connect each model with the universals it accounts for.

p

p

p

p

p

Author's biography

**Eugene V. Koonin** is a Senior Investigator at the National Center for Biotechnology Information (National Library of Medicine, National Institutes of Health) , as well as the editor-in-chief of the journal *Biology Direct*. Dr. Koonin's group performs research in many areas of evolutionary genomics, with a special emphasis on whole-genome approaches to the study of major transitions in life's evolution, such as the origin of eukaryotes, the evolution of eukaryotic gene structure, the origin and evolution of different classes of viruses, and evolutionary systems biology. Dr. Koonin is the author of more than 600 scientific articles and two books: *Sequence - Evolution - Function: Computational Approaches in Comparative Genomics* (with Michael Galperin, 2002) and *The Logic of Chance: The Nature and Origin of Biological Evolution* (2011).

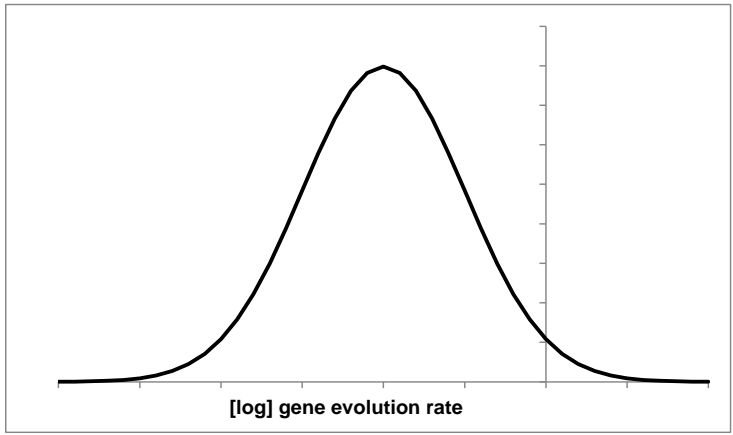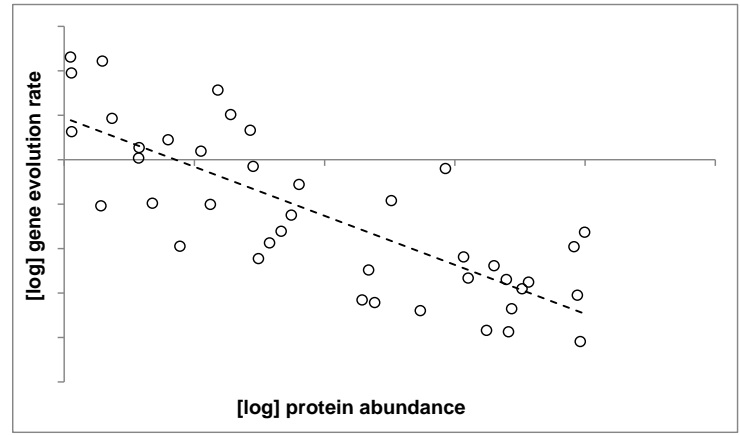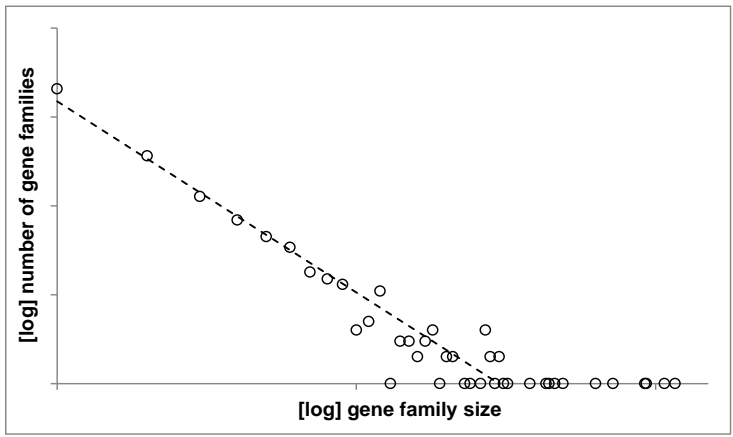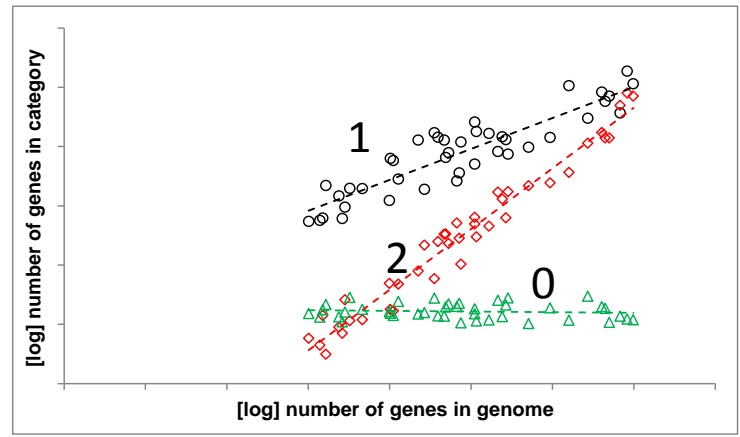

Figure 1

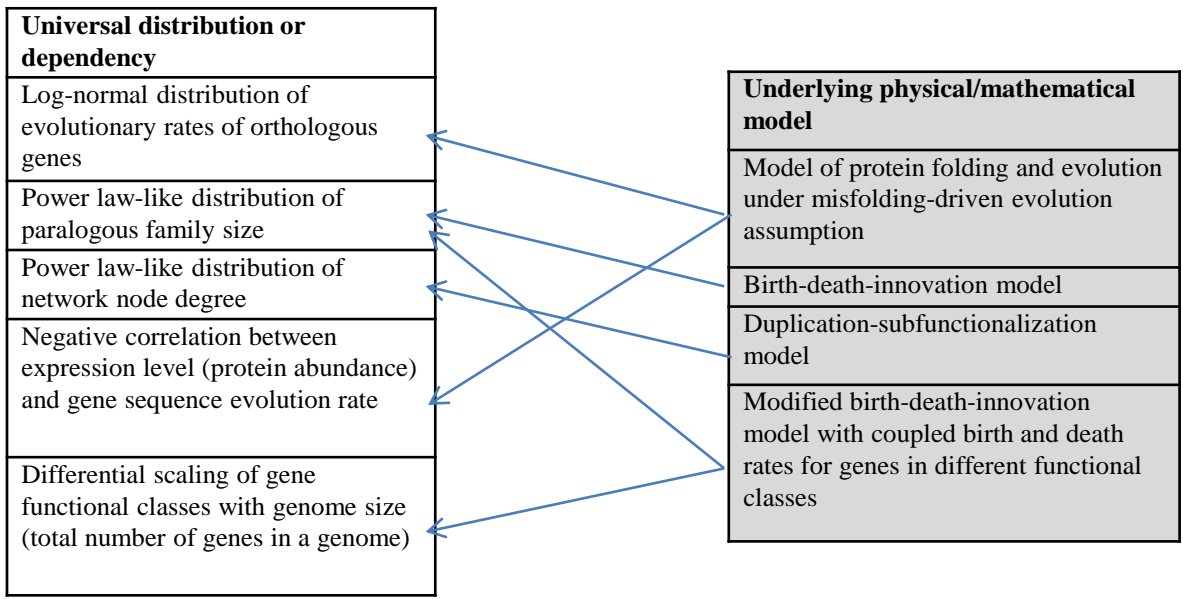

Figure 2